# AI FOR SUSTAINABLE DATA PROTECTION AND FAIR ALGORITHMIC MANAGEMENT IN ENVIRONMENTAL REGULATION

Sahibpreet Singh[*]
Saksham Sharma[**]

## ABSTRACT

*Integration of AI into environmental regulation represents a significant advancement in data management. It offers promising results in both data protection plus algorithmic fairness. This research addresses the critical need for sustainable data protection in the era of ever evolving cyber threats. Traditional encryption methods face limitations in handling the dynamic nature of environmental data. This necessitates the exploration of advanced cryptographic techniques. The objective of this study is to evaluate how AI can enhance these techniques to ensure robust data protection while facilitating fair algorithmic management. The methodology involves a comprehensive review of current advancements in AI-enhanced HE and MPC. It is coupled with an analysis of how these techniques can be applied to environmental data regulation. Key findings indicate that AI-driven dynamic key management, adaptive encryption schemes, and optimized computational efficiency in HE, alongside AI-enhanced protocol optimization and fault mitigation in MPC, significantly improve the security of environmental data processing. These findings highlight a crucial research gap in the intersection of AI, cyber laws, and environmental regulation, particularly in terms of addressing algorithmic bias, transparency, and accountability. The implications of this research underscore the need for stricter cyber laws. Also, the development of comprehensive regulations to safeguard sensitive environmental data. Future efforts should focus on refining AI systems to balance security with privacy and ensuring that regulatory frameworks can adapt to technological advancements. This study provides a foundation for future research aimed at achieving secure sustainable environmental data management through AI innovations.*

## KEYWORDS

Cyber Laws, Sustainable Environmental Data Protection, Secure Multi-Party Computation, Artificial Intelligence (AI), Homomorphic Encryption

---

[*] LLM, Department of Laws, Guru Nanak Dev University, Amritsar, Punjab.
https://orcid.org/0009-0009-9695-2674
[**] LLM, University School of Law, Desh Bhagat University, Fatehgarh Sahib, Punjab.
https://orcid.org/0009-0001-9264-1811

# INTRODUCTION

State of the art tech has become critical in addressing sustainable development concerning the environment. The integration of AI into environmental regulation offers innovative solutions for managing sensitive data. This paper examines the dual role of AI in enhancing sustainable data protection and ensuring fair algorithmic management within the realm of environmental regulation.[1] These regulations necessitate the analysis of vast amounts of sensitive data, for e.g., pollution levels, emission metrics, and ecological monitoring results.[2] Safeguarding this information or protecting this information and simultaneously following the regulations is very important. Although traditional encryption methods are effective, yet they often struggle with the demands of real-time data processing and the evolving nature of cyber threats. HE[3] and Secure MPC[4] are advanced cryptographic techniques that address these challenges. They allow computations on encrypted and collaborative data analysis without exposing sensitive information. HE enables computations to be performed on encrypted data, thus preserving confidentiality during analysis. AI enhances HE through dynamic key management, adaptive encryption schemes, and efficient computation optimizations. For instance, AI-driven dynamic key management predicts potential threats and adjusts key parameters, while adaptive encryption schemes modify HE parameters in response to real-time data access patterns and threats. These enhancements ensure robust data protection while allowing regulatory agencies to perform necessary analyses without compromising sensitive information.[5]

Similarly, MPC facilitates collaborative computations among multiple stakeholders while maintaining data privacy. AI optimizes MPC protocols by predicting efficient communication strategies, detecting faults, and adapting security measures based on real-time threat assessments. This capability is particularly useful in environmental regulation, where data from various entities must be integrated and analyzed securely. AI-enhanced MPC allows for effective joint analyses of environmental data, such as regional air quality assessments, without disclosing proprietary information. The integration of AI with these advanced encryption techniques represents a significant advancement in data protection for environmental regulation. However, AI brings forth legal challenges. Algorithmic bias, black box decision-making, accountability for failures, and compliance with data protection regulations are critical considerations. The reliability factor of these systems must be considered responsibly to address these issues to foster trust.[6] Furthermore, the confluence of cyber laws, environmental programs, and AI offers a holistic approach to managing environmental data and infrastructure. Cyber laws, which govern data security and privacy, intersect with environmental programs that monitor and manage natural resources. The integration of AI enhances this synergy by providing innovative solutions for data analysis, resource management, and environmental monitoring. AI tech, such as ML[7] algorithms for detecting environmental changes and optimizing resource use, contribute to more effective and sustainable environmental practices.[8]

---

[1] Susanne Durst, Christoph Hinteregger & Malgorzata Zieba, *The Effect of Environmental Turbulence on Cyber Security Risk Management and Organizational Resilience*, 137 COMPUTERS & SECURITY 103591 (2024).
[2] P. Asha et al., *IoT Enabled Environmental Toxicology for Air Pollution Monitoring Using AI Techniques*, 205 ENVIRONMENTAL RESEARCH 112574 (2022).
[3] Homomorphic Encryption.
[4] Multi-Party Computation.
[5] Shankha Goswami et al., *The Role of Cyber Security in Advancing Sustainable Digitalization: Opportunities and Challenges*, 3 JOURNAL OF DECISION ANALYTICS AND INTELLIGENT COMPUTING 270 (2023).

## **DATA SECURITY IN CYBER SPACE**

The regulation of environmental has become increasingly data-driven. The need for protection of this data has become very important.[9] This section explores how established cybersecurity frameworks can be adapted to ensure the secure management of data. It will also examine the enforcement of specific cybersecurity standards. Environmental data, including air/water quality, and climate change, is critical for informed decision-making.[10] The value of this data make it a target for cyber-attacks. This can have serious implications for environmental protection efforts and public health. Therefore, implementing robust cybersecurity standards tailored to environmental data is essential.[11]

ISO/IEC 27001 is extensively known for information security management. It provides for a systematic management of sensitive information.[12] It was originally designed for general information security. It can be adapted to meet the specific needs of environmental data management. Its risk based approach is very useful for the following reasons:

---

[6] Sneha Thombre et al., *Prediction of Environmental Impacts Through Artificial Intelligence Techniques*, 23 CAILIAO KEXUE YU GONGYI/MATERIAL SCIENCE AND TECHNOLOGY 406 (2024).
[7] Machine Learning.
[8] SHIVA ROOPAN, INTRO TO AI (2019).
[9] Anna Ubaydullayeva, *Artificial Intelligence and Intellectual Property: Navigating the Complexities of Cyber Law*, 1 INTERNATIONAL JOURNAL OF LAW AND POLICY (2023).
[10] Komang Adi Kurniawan Saputra & Selmita Paranoan, *Do Cyber Security, Digitalisation and Data Visualisation Affect the Quality of Internal Environmental Audits?*, 18 AUSTRALASIAN ACCOUNTING, BUSINESS AND FINANCE JOURNAL 158 (2024).
[11] IMPLEMENTING ENVIRONMENTAL LAW, (Paul Martin & Amanda Kennedy eds., 2015).
[12] Carla Carvalho & Eduardo Marques, *Adapting ISO 27001 to a Public Institution*, 11 *in* 2019 14TH IBERIAN CONFERENCE ON INFORMATION SYSTEMS AND TECHNOLOGIES (CISTI) 1 (2019).

i. **Risk Management**: It underlines the risks associated with information security. This is crucial for environmental data. There is a risk of data breaches to the extent that compliance with relevant laws may be affected.[13]

ii. **Control**: It outlines a set of safeguards designed to protect information assets. These can be tailored to address the specific needs of data management, ensuring secure data collection while protecting data integrity during transmission.[14]

iii. **Continuous Improvement**: The framework advocates for continuous monitoring and improvement of information security practices. This principle supports the dynamic nature of environmental data management, where new threats and vulnerabilities may emerge, requiring ongoing adjustments to security measures.[15]

While ISO/IEC 27001 provides a strong foundation, the unique characteristics of environmental data necessitate the development of specialized standards. Such standards require to cover the following:

a) **Data Classification**: Environmental data varies in sensitivity. It ranges from public data on general air quality to confidential data on specific industrial emissions. Establishing classification schemes can help determine appropriate security measures for different data types.[16]

b) **Governing Amenability**: Environmental data management must comply with various regulations, like GDPR[17] and sector-specific regulations like the Clean Air Act.[18] Specific cybersecurity standards should align with these requirements to ensure comprehensive data protection.

c) **Data Integrity**: Ensuring the accuracy of data is critical for effective regulatory oversight. Standards should include provisions to prevent unauthorized modifications.[19] This should also provide for secure data handling/validation and error-correction.

---

[13] Hamed Taherdoost, *Understanding Cybersecurity Frameworks and Information Security Standards—A Review and Comprehensive Overview*, 11 ELECTRONICS 2181 (2022).
[14] Johnson Odakkal & Neeraj Singh Manhas, *Dynamic Maritime Airspace Management: The Philosophy for an AI Environment*, 216 (2024).
[15] Fereshteh Ghahramani et al., *Continuous Improvement of Information Security Management: An Organisational Learning Perspective*, 32 EUROPEAN JOURNAL OF INFORMATION SYSTEMS 1011 (2023).
[16] CHUNLONG ZHANG, FUNDAMENTALS OF ENVIRONMENTAL SAMPLING AND ANALYSIS (2024).
[17] General Data Protection Regulation.
[18] Tanushree Ganguly, Kurinji L. Selvaraj & Sarath K. Guttikunda, *National Clean Air Programme (NCAP) for Indian Cities: Review and Outlook of Clean Air Action Plans*, 8 ATMOSPHERIC ENVIRONMENT: X 100096 (2020).
[19] Mahkamov Durbek, *The Role of Cybersecurity in Environmental Law: Ensuring the Protection of Sensitive Data in the Age of Digitization*, in LEGAL TECH, EDUCATION AND DIGITAL TRANSFORMATION OF LAW 80 (2023).

d) **Incident Response and Recovery**: Standardizing incident response is essential. This is important to deter potential cyber-attacks. This includes protocols that try to lessen the effects of security related incidents.[20] It also involves for data breach notifications, forensic investigations, and recovery processes.

The following examples seek to elaborate on the application of these standards:

1. **Smart Grid Technologies**: These are increasingly used to manage environmental parameters. It helps monitor energy consumption and emissions. Applying cybersecurity protocols ensures that the data collected is protected from any tampering. This helps maintain the accuracy of environmental monitoring systems.[21]

2. **Climate Data Management**: Adapting ISO/IEC 27001 by organizations managing large volumes of climate data addresses data storage and transmission security. It can help safeguard the data against potential cyber threats. This ensures that the data remains reliable for climate modeling.[22]

Cyber laws vary a lot from one country to another, which reflects different legal traditions, cultural values and technological advancement levels. The transnational nature of internet demands some international cooperation and harmonization of laws. The United Nations and International Telecommunication Union are among the major international organizations which facilitate global cooperation and worldwide rules on cyberspace.

a) **European Union**

The EU is a prominent region in making variety of cyber laws especially concerning to data protection and cybersecurity. It has the strictest data protection laws; GDPR which was implemented in 2018 is known to have one of the most stringent privacy and data security standards internationally. Moreover, EU also adopted NIS Directive to secure crucial services and providers of digital services against cyber-attacks.[23]

b) **United States**

US internet law can be said to consist of both sets of laws at the federal and state levels, which is quite normal. The US Federal Trade Commission is basically responsible for implementing the data protection and privacy laws while the Department of Homeland

---

[20] Christos Beretas, *Information Systems Security, Detection and Recovery from Cyber Attacks*, Volume 1 UNIVERSAL LIBRARY OF ENGINEERING TECHNOLOGY 27 (2024).
[21] James Henry & Murad Aziz, *Cybersecurity Challenges in Sustainable Data Management: Addressing Threats to Environmental Responsibility*, (2024).
[22] Alan Calder & Steve Watkins, *IT Governance: An International Guide to Data Security and ISO 27001/ISO 27002*, 1 (2024).
[23] Gayatri Gaddamanugu & Kamaneeya Paku, *New Insight into CO2 Sequestration: A Blended Approach of Artificial Intelligence and Machine Learning Tools*, 962 KEY ENGINEERING MATERIALS 139 (2023).

Security focuses on cybersecurity related concepts. To enhance cyber security, particularly with respect to digital wellbeing, the US focuses on public-private partnerships and recognizes that the private sector plays a crucial role.

### c) Asia-Pacific

Countries in Asia-Pacific have also shown interest in coming up with cyber laws. For example, Japan has laws relating to personal information known as Act on Protection of Personal Information. Cybersecurity law of the People's Republic of China enacted by China also provides for data location as well as network operator's cybersecurity requirements.

**Challenges and Future Measures**

Although many improvements have been made, there are still some weak spots in the field of cyber laws. The most significant of these challenges is how well we can keep pace with fast changing technology. As technology evolves, so do the techniques that are employed by hackers hence necessitating constant upgrading of legal frameworks. Additionally, the worldwide nature of cyber threats means that there has to be increased international cooperation as well as harmonization of laws which may be difficult due to different national interests and legal systems. The other one could be seen as finding a middle ground between privacy and security. While strict measures should be implemented to curb the threats posed on this platform, at times people's rights to privacy may be infringed upon through strong cybersecurity policies. This is a difficult balance to strike and must involve careful thinking and an ongoing conversation among multiple parties involved.[24] In future, development of cyber laws will probably focus around several areas such as:

1. **AI and Cybersecurity**- The mixing of AI into cybersecurity practices may lead to new legal and ethical challenges. It is crucial that AI-powered cyber security tools are transparent, responsible and without any bias.[25]

2. **Data Individuality and Localization**- Countries may adopt more stringent data localization requirements in response to issues about data sovereignty, which could affect global data streams as well as digital commerce. Exploring these laws is important for companies operating across borders.

3. **Digital Identification and Blockchain**- Advancements in digital identity and blockchain technology have the potential to revolutionize how personal information is taken care of and protected. In addition, it will be necessary to establish legal

---
[24] John Owen, *Big Data and AI in IoT: Opportunities and Challenges* (2024).
[25] Hwee-Joo Kam, Chen Zhong & Allen Johnston, *The Impacts Of Generative AI on the Cybersecurity Landscape* (2024).

guidelines for supporting these advances while protecting individual privacy and security.[26]

In the digital age, the era of cyber laws has become an inevitable fact; these are frameworks required for defense against internet crimes on individuals, organizations and even governments. Cyber laws need to keep up with changing technology to cater for different requirements according to various opportunities and challenges.[27] This paper looks at how cyber laws have developed over time as well as their extension. It also tries to address future prospects using international perspectives so that a reliable and sustainable digital future can be envisaged.

## CONCERNS RELATING ENVIRONMENT

Human activities worsen the plethora of problems faced by the environment. Global warming which is directly caused by climate change and is primarily fueled by release of greenhouse gases. These have very bad effects on environment and human civilization. Some examples of these include increased temperatures, ice caps melting as well as more frequent extreme weather conditions. Pollution is another major problem that affects air quality, water bodies' status and soil conditions, thus posing danger to mankind and animals.

**Impact of Technology on the Environment**

Technology has a twofold impact on the environment. On one hand, technological advancements have contributed to environmental declination.[28] The artificial revolution, powered by fossil energies, has led to increased carbon emigrations and pollution. The production and disposal of significant quantities of e-waste has detrimental effects on the soil and water. Also, the growing demand for data centers results in substantial energy consumption and carbon emigrations.[29]

On the other hand, technology also offers results to environmental problems. Advances in renewable energy technologies, similar as solar, wind, and hydroelectric power, have the eventuality to reduce our reliance on fossil energies and drop greenhouse gas emissions for the time being. Smart grid technology improves the effectiveness of electricity distribution, farther supporting the integration of renewable energy sources. Innovations in battery storage and electric vehicles contribute to cleaner transportation systems, reducing air pollution and dependence on oil in the short term.

---

[26] Анастасия Александровна Сеньковская & Марат Урасканович Байдельдинов, *Digital Identification Technologies Based on Blockchain* 47 (2024).
[27] Sheriffdeen Kayode, *Navigating Regulatory and Legal Challenges in AI and ML-Powered Cybersecurity: A Comprehensive Analysis* (2023).
[28] Eleonora Viganò, Michele Loi & Emad Yaghmaei, *Cybersecurity of Critical Infrastructure*, in THE INTERNATIONAL LIBRARY OF ETHICS, LAW AND TECHNOLOGY 157 (2020).
[29] Payal Kathuria & Swapnil Aggarwal, *Impact of Technology on Environment: A Review of Literature*, 11 JOURNAL OF INTERNATIONAL ACADEMIC RESEARCH FOR MULTIDISCIPLINARY 2320 (2023).

**Environmental Monitoring and Operation**

Technology is a very considerable factor in Environmental monitoring & operation. Remote seeing technologies, like satellites and drones, give precious data for tracking changes in land use, timber cover, and water management. These tools help scientists and policymakers understand the impacts of mortal conditioning on the terrain and develop strategies for conservation and sustainable resource operation.

GI[30] Systems (Civilians) assist in the collection and interpretation of spatial data for environmental management. For illustration, it can be used to identify areas at threat of natural disasters, plan protected areas, and cover wildlife territories. Environmental detectors and IoT[31] collect real-time data on air and water quality, soil humidity, and other environmental parameters, enabling further effective monitoring and response to environmental hazards.[32]

**Regulations and Programs**

Governments have enforced a range of programs aimed at addressing environmental challenges. These measures are designed to focus on sustainable development.

1. **Transnational Agreements**

Transnational agreements play a pivotal part in addressing global environmental issues. The Paris Agreement is a corner accord within the UNFCCC[33]. It objects to limit global temperature below 2 degrees Celsius of pre-industrial levels. Further, attempts are made to make sure that the temperature does not rise more than 1.5 degrees Celsius. Signatory countries are needed to submit NDCs[34] outlining their climate conduct and regularly report on their progress.[35] CBD[36] is another crucial transnational agreement aimed at conserving natural diversity, promoting sustainable development. The CBD sets global targets for biodiversity conservation and encourages countries to develop public biodiversity strategies and action plans.[37]

---

[30] Geographic Information.
[31] Internet of Things.
[32] Dwijendra Dwivedi, Ghanashyama Mahanty & Varunendra Dwivedi, *Intelligent Conservation: A Comprehensive Study on AI-Enhanced Environmental Monitoring and Preservation* 215 (2024).
[33] United Nations Framework Convention on Climate Change.
[34] Nationally Determined Contributions.
[35] Thomas Hickmann et al., *The United Nations Framework Convention on Climate Change Secretariat as an Orchestrator in Global Climate Policymaking*, 87 INTERNATIONAL REVIEW OF ADMINISTRATIVE SCIENCES 21 (2021).
[36] Convention on Biological Diversity.

### 2. National Environmental Regulations

Countries have legislated different public environmental regulations to address specific environmental issues. For example, the States has provisions dedicated to either separate Air Quality norms (like the Clean Air Act) or Water Quality, (like the Clean Water Act). These laws are executed by EPA[38]. These have contributed to significant advancements in air and water quality over many decades.[39] The EU has enforced the EIA[40] Directive.[41] The EU's ETS[42] is a crucial policy tool for reducing greenhouse gas by setting a cap on emission. This, however, permits companies to purchase and sell quota for emissions.[43]

### 3. Commercial Environmental Responsibility

Commercial environmental responsibility involves reducing the environmental impact of business operations/products/services. This includes achieving energy efficiency and reducing waste sustainably. Companies may also engage in environmental conditioning by adopting reforestation policies.[44] The conception of indirect frugality is gaining traction as a sustainable business model. It seeks to reduce waste and maximize the life utilization of the product by extending product lifecycle through recycling, reuse and refurbishing. AI tech can enhance our capability to address environmental challenges.[45]

### 4. Environmental Monitoring

AI-powered systems can dissect vast quantities of environmental data to identify patterns and prognosticate unborn trends. ML algorithms can reuse data from remote sensing technologies to track deforestation, wildlife populations, and assess the health of ecosystems. AI can also ameliorate climate modeling by integrating different data sources and refining prognostications of rainfall patterns, ocean-level rise, and other climate-related phenomena.[46]

---

[37] Felicity Keiper & Ana Atanassova, *Regulation of Synthetic Biology: Developments Under the Convention on Biological Diversity and Its Protocols*, 8 FRONT. BIOENG. BIOTECHNOL. (2020).
[38] Environmental Protection Agency.
[39] ROBERT V. PERCIVAL ET AL., ENVIRONMENTAL REGULATION: LAW, SCIENCE, AND POLICY (2021).
[40] Environmental Impact Assessment.
[41] PADC UNIT, ENVIRONMENTAL IMPACT ASSESSMENT (1983).
[42] Emigrations Trading System.
[43] Patrick Bayer & Michaël Aklin, *The European Union Emissions Trading System Reduced $CO_2$ Emissions despite Low Prices*, 117 PROC. NATL. ACAD. SCI. U.S.A. 8804 (2020).
[44] LUCAS BERGKAMP, LIABILITY AND ENVIRONMENT: PRIVATE AND PUBLIC LAW ASPECTS OF CIVIL LIABILITY FOR ENVIRONMENTAL HARM IN AN INTERNATIONAL CONTEXT (2021).
[45] Kannan Govindan, *How Artificial Intelligence Drives Sustainable Frugal Innovation: A Multitheoretical Perspective*, 71 IEEE TRANS. ENG. MANAGE. 638 (2024).
[46] Rishaba Mohita et al., *Sustainability Development Metric Validation of Indian Territory Using Regression Approach* (2024).

5. **Energy Efficiency and Management**

Energy consumption can be managed and energy systems can be improved with the help of AI. For e.g., AI algorithms can analyze energy usage patterns in buildings and industrial facilities to identify prospects for energy savings. AI-equipped smart grids can balance supply and demand, integrate renewable energy sources, and reduce energy annihilation. AI may also increase the performance of renewable energy systems by regulating the operation of solar panels and wind turbines efficiently.[47]

6. **Sustainable husbandry**

AI technologies can support sustainable husbandry practices[48] by optimizing the use of coffers similar as water, diseases, and fungicides. Precision husbandry ways, powered by AI, enable growers to cover soil conditions, crop health, and rainfall patterns in real- time. This allows for targeted interventions that minimize resource use and reduce environmental impacts. AI can also ameliorate force chain operation, reducing food waste and icing that agrarian products reach consumers efficiently.

7. **Waste Management and Recycling**

AI can enhance waste recycling processes by perfecting sorting and recovering operations. AI-powered robots can sort recyclables more efficiently than manual workers[49]. ML algorithms can dissect data on waste generation and disposal to identify trends and optimize waste collection routes. AI can also support the development of indirect efficiency models by means of reusing and repurposing materials.

## AUGMENTING ENVIRONMENTAL DATA PROTECTION

The integration of AI in encryption techniques presents a promising avenue for achieving sustainable data protection while facilitating regulatory compliance. This section explores advanced encryption methods to manage environmental data effectively.[50]

**Homomorphic Encryption and AI Integration**

HE[51] is a cryptographic technique. This enables such computations directly on encrypted data. There is no prerequisite to perform the inverse operation. This feature is particularly valuable in environmental regulation as sensitive data needs to be analyzed without being exposing. Following are the AI-Enhanced HE Techniques:

   i. **Dynamic Key Management**: AI can optimize the management of encryption keys by predicting potential threats and adjusting key parameters dynamically. ML models can analyze historical data to foresee vulnerabilities, thus enhancing the security of the encrypted data.[52]

   ii. **Adaptive Encryption Schemes**: AI algorithms can be employed to adjust the parameters of HE schemes in real-time based on the data being processed and the current threat landscape. For instance, an AI system could modify the depth of polynomial approximations or the encryption parameters based on detected anomalies or changes in data access patterns.[53]

   iii. **Efficient Homomorphic Computation**: AI can optimize the computational efficiency of HE operations. By using machine learning techniques to predict the most efficient encryption and decryption paths or by pre-computing certain values, AI can significantly reduce the computational overhead associated with HE.[54]

---

[47] Cheng Chen et al., *Artificial Intelligence on Economic Evaluation of Energy Efficiency and Renewable Energy Technologies*, 47 SUSTAINABLE ENERGY TECHNOLOGIES AND ASSESSMENTS 101358 (2021).
[48] SHIV SINGH, *Instrumentation, Technologies and Application of Sensors in Animal Husbandry Mechanization*, (2023).
[49] Jayaram M.A, *AI Manifolds in Waste Management*, in WASTE OR WEALTH (WoW), 3 DAYS WORKSHOP (2024).
[50] U. S. Gahtan, *Encrypted AI for Environmental Monitoring Systems*, 3 INTERNATIONAL JOURNAL OF RESEARCH RADICALS IN MULTIDISCIPLINARY FIELDS, 24 (2024).
[51] Homomorphic Encryption.

In environmental regulation, HE could be utilized to protect data from sensors and monitoring systems. For instance, a regulatory agency could use HE to analyze encrypted data from multiple sources without ever accessing the raw data. This ensures that sensitive information remains confidential while still allowing for regulatory compliance and analysis.

**Secure Multi-Party AI-Enhanced Computation**

Secure MPC[55] is another advanced cryptographic technique. It allows multiple parties to collaboratively compute a function over their inputs while keeping those inputs private. This is useful in scenarios where data from various stakeholders need to be analyzed without revealing individual datasets. Here some of the techniques are:

   a) **Optimized Protocols**: AI can enhance the efficiency of MPC protocols by learning from previous computations to optimize the communication and computation strategies among parties. ML models can predict the most efficient configurations based on the nature of the data plus the function to be computed.[56]

---

[52] Shahnawaz Ahmad et al., *Machine Learning-Based Intelligent Security Framework for Secure Cloud Key Management*, 27 CLUSTER COMPUT 5953 (2024).
[53] Leyli Karaçay, Erkay Savaş & Halit Alptekin, *Intrusion Detection Over Encrypted Network Data*, 63 THE COMPUTER JOURNAL 604 (2020).
[54] Ardhi Wiratama Baskara Yudha et al., *BoostCom: Towards Efficient Universal Fully Homomorphic Encryption by Boosting the Word-Wise Comparisons*, (2024).
[55] Multi-Party Computation.
[56] Ian Zhou et al., *Secure Multi-Party Computation for Machine Learning: A Survey*, 12 IEEE ACCESS 53881 (2024).

b) **Fault Mitigation**: AI can be used to mitigate anomalies in MPC processes. By continuously monitoring the computation processes, AI systems can identify potential security breaches; then take corrective actions to ensure the integrity of the results.[57]

   c) **Adaptive Security Measures**: Similar to HE, AI can help in dynamically adapting security measures in MPC based on real-time threat assessments. For example, AI can adjust the level of data sharing or encryption strength depending on detected risk factors or changes in the regulatory environment.[58]

In environmental regulation, MPC can be employed to allow multiple stakeholders to contribute to a joint analysis of environmental data without compromising their proprietary information.[59] For instance, different organizations could collaborate to assess regional air quality or climate impact without disclosing their individual datasets. AI-enhanced MPC would ensure that the process is both secure and efficient, accommodating changes in threat landscapes and data sharing requirements.

**Implementation in Environmental Data Regulation**

**Scenario: Monitoring and Reporting**

Consider a scenario where multiple companies and regulatory agencies need to monitor and report on pollution levels from various sources.[60] Using HE, the data collected from different sensors can be encrypted, and the HE scheme can be dynamically adjusted by AI to handle varying data types and threats. Meanwhile, MPC can be employed to allow these entities to jointly compute aggregate pollution metrics while keeping individual data confidential. AI systems can ensure that the encryption and computation processes are optimized and secure, handling potential threats and inefficiencies.

**Outcome**

The integration of AI with HE and MPC provides a robust framework for protecting environmental data while allowing necessary computations and regulatory analysis. By dynamically adapting encryption techniques to emerging threats and optimizing computational efficiency, AI enhances the sustainability of data protection measures in environmental regulation.

---

[57] M. B. Farbman, *Encrypted AI Techniques for Anomaly Detection*, 3 INTERNATIONAL JOURNAL OF RESEARCH AND REVIEW TECHNIQUES 76 (2024).
[58] Yasser D. Al-Otaibi, *Distributed Multi-Party Security Computation Framework for Heterogeneous Internet of Things (IoT) Devices*, 25 SOFT COMPUT 12131 (2021).
[59] Anil Kumar Yadav Yanamala & Srikanth Suryadevara, *Navigating Data Protection Challenges in the Era of Artificial Intelligence: A Comprehensive Review*, 15 REVISTA DE INTELIGENCIA ARTIFICIAL EN MEDICINA 113 (2024).
[60] Megha Chauhan & Deepali Sahoo, *Towards a Greener Tomorrow: Exploring the Potential of AI, Blockchain, and IoT in Sustainable Development*, 23 NATURE ENVIRONMENT AND POLLUTION TECHNOLOGY 1105 (2024).

The fusion of AI with Homomorphic Encryption and Secure Multi-Party Computation represents a significant advancement in data protection for environmental regulation. AI-driven enhancements can provide dynamic, efficient, and secure solutions to handle sensitive environmental data, ensuring that regulatory needs are met without compromising data privacy. This approach not only addresses current challenges in data protection but also sets the stage for future innovations in secure and fair environmental management.[61]

**Ethical and Legal Considerations**

AI brings about significant ethical and legal considerations. AI systems must are developed and used responsibly to address bias, translucency, responsibility, and sequestration[62].

i. **Bias and Fairness**

It is important to make sure that the AI systems that are used for encryption and data management are not biased or further increase any existing bias. This would involve not only addressing biases in training data but also ensuring that encryption methods and MPC protocols are designed and tested to avoid unfair outcomes. In the context of encryption, bias might affect how encryption algorithms are designed or how access to encrypted data is managed. Ensuring fairness involves ongoing evaluation of these systems to detect and correct any biases.[63]

ii. **Translucency and Explainability**

Understanding how these advanced encryption techniques work is crucial for compliance. Explainable AI approaches can be used to enhance transparency in encryption methods. This can ensure stakeholders understand how data is protected plus how secure computations are performed. This is particularly relevant in regulatory environments where decisions based on data need to be accountable.[64]

iii. **Liability**

Determining who is accountable for failures in encryption/inaccuracies in data handling is essential for maintaining regulatory compliance and trust. In the case of data breaches or errors in encrypted data handling, clear legal guidelines are needed to determine responsibility.[65] This ensures that all parties involved in the implementation of AI-driven encryption systems are accountable.

---

[61] Ha Eun David Kang et al., *Homomorphic Encryption as a Secure PHM Outsourcing Solution for Small and Medium Manufacturing Enterprise*, 61 JOURNAL OF MANUFACTURING SYSTEMS 856 (2021).
[62] Edwin Frank, *Ethical Considerations in Artificial Intelligence (AI)* (2024).
[63] Divya Dwivedi, *Algorithmic Bias: A Challenge for Ethical Artificial Intelligence (AI)* 65 (2023).
[64] Yi-Fan Wang et al., *Citizens' Trust in AI-Enabled Government Systems*, INFORMATION POLITY 1 (2024).
[65] JOSHUA C. GELLERS, RIGHTS FOR ROBOTS: ARTIFICIAL INTELLIGENCE, ANIMAL AND ENVIRONMENTAL LAW (EDITION 1) (2020).

### iv. Sequestration

The usefulness of AI systems in particular, makes it essential to go the extra mile to ensure that data remains properly secured. AI-powered encryption methods, including HE and MPC, are designed with privacy in mind. Ensuring that these techniques are implemented in a way that respects privacy rights is crucial for their effective use in environmental regulation.[66]

## Legal fabrics Governing AI

AI development and usage are also controlled by governments and other international organizations.[67] The aim is to maintain public trust associated with AI while promoting innovation.

### a. European Union[68]

The European Union has taken a visionary approach to AI regulation. The European Commission's Artificial Intelligence Act aims to produce a comprehensive nonsupervisory frame for AI, fastening on high-threat AI operations. The legislation includes conditions for translucency, responsibility, and human oversight, as well as vittles for addressing bias.

### b. United States

In the United States, civil and state agencies address different aspects of AI. FTC[69] enforces regulations related to data sequestration. While NIST[70] develops norms for AI.[71] Efforts are underway to develop a more coordinated approach to AI regulation in order to regulate AI governance.

### c. International Associations

Transnational associations, like the United Nations and the Organization for OECD[72], are also playing a part in developing global morals and norms for AI.[73] The OECD's AI Principles, espoused by over 40 countries, give guidelines for the responsible development and use of AI.

## CONCLUSION AND SUGGESTIONS

The paper has explored the intricate corners of cyber laws, environmental programs, and AI. Integration of AI into environmental programs, supported by robust cyber laws, offers transformative results. AI's ability to analyze large amounts of data can support operation and monitoring of the environment in a big way. For e.g., AI-driven systems can prognosticate climate change patterns, optimize renewable energy use, and improve waste management. These not only contribute to sustainability but also promote the effective use of resources, aligning with global environmental objectives. The integration of AI further benefits from advanced encryption techniques. HE allows operations to be done on encrypted data without de-cryption. Thus, safeguarding sensitive environmental data while facilitating in-depth analysis. AI enhances HE through dynamic key management, adaptive encryption schemes, and optimized homomorphic computations, thereby fortifying data security and analysis. In the same way, MPC enables various stakeholders to jointly compute functions based on their private inputs while ensuring the confidentiality of the data. AI contributes to MPC by optimizing protocol efficiency, detecting faults, and adapting security measures in real-time, ensuring secure and effective multi-party collaborations.

---

[66] *Id.*
[67] Amaka Obinna & Azeez Kess-Momoh, *Developing a Conceptual Technical Framework for Ethical AI in Procurement with Emphasis on Legal Oversight*, 19 GSC ADVANCED RESEARCH AND REVIEWS 146 (2024).
[68] Serena Quattrocolo & Ernestina Sacchetto, *EU and AI* 636 (2024).
[69] Federal Trade Commission.
[70] National Institute of norms and Technology.
[71] LUCA NANNINI, AGATHE BALAYN & ADAM SMITH, EXPLAINABILITY IN AI POLICIES: A CRITICAL REVIEW OF COMMUNICATIONS, REPORTS, REGULATIONS, AND STANDARDS IN THE EU, US, AND UK (2023).
[72] Economic Co-operation and Development.
[73] Oluomachukwu Chilaka, *The Role of International Conferences in Advancing AI Education* (2024).

The deployment of AI in environmental domain necessitates strict cyber laws to protect data sequestration. GDPR play a pivotal part in securing environmental data. Ensuring non-discriminatory AI operations is essential to ensure equitable public trust. Despite the promising prospects, several challenges must be addressed to harness the full eventuality of AI, cyber laws, and environmental programs. Data security remains primary concern; especially, AI systems use large datasets to work effectively. This can lead to serious security concerns. Meeting data protection requirements is necessary for preventing people from spilling the confidential environmental data. The black-box nature of AI is another major complicating factor. Developing unprejudiced datasets and incorporating fairness into algorithms are critical way in this direction. Policymakers must collaborate with other stakeholders to produce comprehensive and flexible regulations that address the evolving landscape of AI. Looking forward, the confluence of AI, cyber laws, and environmental programs holds immense eventuality for innovation and sustainability. Enhanced environmental monitoring, bettered resource operation, and optimized energy use are just a many areas where AI can make a significant impact. Thus, through innovation we can pave the way for a more sustainable and secure future, benefiting our environment. By implementing these targeted suggestions, stakeholders can significantly enhance the efficiency of environmental data management, ensuring robust protection against cyber threats while facilitating compliance with regulatory requirements:

1. **Development of AI-Enhanced Homomorphic Encryption Frameworks-** Research should focus on creating specialized HE frameworks tailored for environmental data processing. Implement adaptive polynomial approximation techniques that can dynamically adjust based on real-time data access patterns. This can reduce computational overhead while maintaining encryption integrity. Incorporate ML models for predicting optimal parameter settings based on historical access data to enhance efficiency and security.

2. **Robust Dynamic Key Management Systems-** Integrate AI-driven dynamic key management systems that utilize predictive analytics to identify potential vulnerabilities. These systems should analyze environmental data access patterns to adjust key lengths and rotation frequencies automatically. Employ anomaly detection algorithms to trigger immediate re-keying processes in response to detected threats.

3. **Real-Time Adaptive Encryption Schemes-** Design adaptive encryption schemes capable of modifying encryption parameters in real-time. These schemes must incorporate a feedback loop where AI algorithms continuously assess current threat levels and operational requirements, allowing for immediate adjustments to encryption methods. This will protect the sensitive environmental data remains from ever evolving cyber threats.

4. **Optimized Secure MPC Protocols-** Improve existing MPC protocols by integrating AI to streamline communication and computational strategies. Develop ML algorithms that analyze previous MPC processes to inform the configuration of future collaborations, optimizing both security and efficiency. Establish criteria for adaptive security measures that respond to real-time threat assessments. This will allow stakeholders to adjust data-sharing protocols accordingly.

5. **Automated Fault Detection and Mitigation Systems-** Implement AI systems that continuously monitor the integrity of MPC processes. Develop algorithms capable of identifying anomalies in real-time, triggering automated fault mitigation measures. This includes rerouting computations or temporarily suspending access to sensitive data until the integrity of the data and processes can be verified.

6. **Comprehensive Data Taxonomy Structures-** Create advanced taxonomy systems specifically for environmental data, categorizing it based on sensitivity and regulatory requirements. This taxonomy should inform the application of encryption methods and data handling procedures, ensuring that more sensitive data receives heightened protection and compliance scrutiny.

7. **Enhanced Regulatory Compliance Tools-** Develop AI-powered tools that automate compliance monitoring for environmental regulations like GDPR and the Clean Air Act.

These tools should be capable of conducting real-time audits of data handling practices. They must also flag any potential non-compliance issue. This will proactively help address compliance risks before they result in legal repercussions.

8. **Ethical AI Development Practices-** Establish frameworks for the ethical development of AI systems used in environmental data protection. This should include rigorous bias detection/mitigation protocols during the AI training phases. Implement transparent mechanisms that help trackback the decision making process. Making this particular regarding data access and processing.

9. **Transnational Standards for AI in Environmental Regulation-** Advocate for the establishment of international standards governing the use of AI in environmental regulation. These standards should focus on the interoperability of AI systems. They must ensure consistent data protection practices across borders. Engage with organizations like the OECD to promote collaborative efforts in developing these guidelines.

10. **Future-Proof Cybersecurity Measures-** Encourage ongoing research into future-proof cybersecurity measures that can adapt to technological advancements. This includes exploring quantum-resistant encryption methods that can withstand emerging threats. Establish partnerships with cybersecurity firms to integrate cutting-edge solutions into existing data protection frameworks.